


\def\BulletItem #1 {\item{$\bullet$}{#1}}


\def\section #1    {\bigskip\noindent{\bf  #1 }\par\nobreak\smallskip}

\def\subsection #1 {\medskip\noindent{\it [ #1 ]}\par\nobreak\smallskip}

\def\eprint#1{$\langle$e-print archive: #1$\rangle$}



%
 \let\miguu=\footnote
 \def\footnote#1#2{{$\,$\parindent=9pt\baselineskip=13pt%
 \miguu{#1}{#2\vskip -7truept}}}
%
%


\def\=>{\Rightarrow}
\def\==>{\Longrightarrow}
 
 \def\dal{\displaystyle{{\hbox to 0pt{$\sqcup$\hss}}\sqcap}}
 
%
\def\lto{\mathop
        {\hbox{${\lower3.8pt\hbox{$<$}}\atop{\raise0.2pt\hbox{$\sim$}}$}}}
\def\gto{\mathop
        {\hbox{${\lower3.8pt\hbox{$>$}}\atop{\raise0.2pt\hbox{$\sim$}}$}}}
%
 




\def\to{\rightarrow}		




\def\Reals{{\rm I\!\rm R}}	

\def				
  \Complexes
   {{\rm C}\llap{\vrule height6.3pt width1pt depth-.4pt\phantom t}}




\def\interior #1 {  \buildrel\circ\over  #1}     



%
%



\def \singlespace{\baselineskip=12pt}  
\def \sesquispace{\baselineskip=16pt}  

\magnification=\magstep1 
\raggedbottom
\hsize=6 true in
 \hoffset=0.27 true in
\vsize=8.5 true in
 \voffset=0.28 true in

\overfullrule=0pt 

\sesquispace  


\rightline{SU-GP-93-12-2}
\rightline{gr-qc/9706002}

\bigskip

\bigskip
\centerline{{\bf FORKS IN THE ROAD, ON THE WAY TO QUANTUM GRAVITY}\footnote{*}
{{\noindent\parindent=0pt\baselineskip=12pt
This paper is the text of a talk given at the symposium on {\it Directions
in General Relativity}, held at the University of Maryland, College Park in
May, 1993, in honor of Dieter Brill and Charles Misner.  To be published in
{\it Int. J. Mod. Phys. A}.
}}}
\bigskip
\baselineskip=12pt  
\centerline {\it Rafael D. Sorkin}
\smallskip
\centerline {\it Instituto de Ciencias Nucleares, 
                 UNAM, A. Postal 70-543,
                 D.F. 04510, Mexico}
 

\smallskip
\centerline { \it and }
\smallskip
\centerline {\it Department of Physics, 
                 Syracuse University, 
                 Syracuse, NY 13244-1130, U.S.A.}

\smallskip
\centerline {\it \qquad\qquad internet address: sorkin@xochitl.nuclecu.unam.mx}
 
\bigskip
\centerline{\it Abstract}
\smallskip
\leftskip=1.5cm\rightskip=1.5cm \singlespace
In seeking to arrive at a theory of ``quantum gravity'', one faces
several choices among alternative approaches.  I list some of these
``forks in the road'' and offer reasons for taking one alternative
over the other.  In particular, I advocate the following: the
sum-over-histories framework for quantum dynamics over the
``observable and state-vector'' framework; relative probabilities over
absolute ones; spacetime over space as the gravitational ``substance''
(4 over 3+1); a Lorentzian metric over a Riemannian
(``Euclidean'') one; a dynamical topology over an absolute one;
degenerate metrics over closed timelike curves to mediate
topology-change; ``unimodular gravity'' over the unrestricted
functional integral; and taking a discrete underlying structure (the
causal set) rather than the differentiable manifold as the basis of
the theory.  

In connection with these choices, I also mention some
results from unimodular quantum cosmology, 
sketch an account of the origin of black hole entropy, 
summarize an argument that the quantum mechanical measurement scheme
breaks down for quantum field theory,
and offer a reason why the cosmological constant of the
present epoch might have a magnitude of around $10^{-120}$ in natural
units. 
\bigskip
\leftskip=0truecm\rightskip=0truecm         
\sesquispace                                
\bigskip\medskip
 

\vfill\break

\noindent{\bf I. A laundry list of alternatives concerning Quantum Gravity}
\medskip
\nobreak     
The organizers have named this conference ``Directions in General
Relativity'', and in that spirit I want to talk in some generality about
directions in Quantum Gravity.  On the way to a theory of quantum gravity
there are many forks in the road, or in other words alternatives one must
choose among, or questions one must answer, before proceeding farther.  I
will begin by listing some of those alternatives and questions which seem
to me the most important, and then I will advocate answers in a manner that
tries to place the choices in an overall context.  I hope that along the
way, a coherent approach to quantum gravity will be seen to emerge.

Also along the way, I will mention a few new or lesser-known results
relevant to the alternatives we will be considering, including an
interpretation of black hole entropy and a possible ``non-unitarity''
associated with ``unimodular'' quantum cosmology.  But first the laundry
list itself (including a personal selection of references to represent the
alternatives\footnote{*}
{These references, like those in the rest of this paper, are meant to be
indicative rather than comprehensive.})

\item{} Is the signature of the spacetime metric {\it Lorentzian\/} [1]]
or {\it Euclidean} [2] (or both [3])? 
     
\item{} Should we allow {\it degeneracies in the metric} [4]  or {\it closed
timelike curves\/} [5]  (or possibly both)?
     
\item{} Should we {\it fix the 4-volume} in the gravitational
Sum-over-histories [6] or extend the  {\it sum over all 4-volumes}, 
as is normally assumed [7]? 

\item{} Is the deep structure of spacetime {\it discrete\/} [8] or {\it
 continuous\/} [8]? 
     
\item{} Which feature of spacetime is most basic, 
     its {\it causal order\/} [9], 
     its {\it metric\/} [10], 
  or its {\it topology\/} [11]
  (or perhaps even the {\it algebra\/} [12] of functions on spacetime)? 
     
\item{} Is topology {\it dynamical} [13] or is it {\it absolute} [14]?
     
\item{} Is the entropy of a black hole {\it outside} [15] or 
 {\it inside} [16] its horizon?
       
\item{} What really exists, the {\it history} [17] or the 
                          {\it wave-function} [18]? 
     
\item{}    Should we approach the ``quantization'' of gravity via the {\it
 sum-over-histories} [19] or via {\it canonical quantization} [20]?
     
\item{}  Is probability  {\it absolute/unconditional} [21] or 
        {\it relative/conditional} [22]?  
        \ \ (do quantum probabilities make sense?)
     
\item{} Is the cosmological constant, $\Lambda$, {\it approximately} [23]
  or {\it exactly} [24] zero?

\bigskip     
 
{\noindent\bf II. What is {\underbar{Classical}} Gravity?}
\smallskip\nobreak     

In order to begin placing these alternatives in context, it is useful to go
back to the classical theory we are trying to ``quantize''.  Like the
majority of physical theories, General Relativity has a threefold structure
comprising a ``kinematical (or substantial) part'', answering the question
``{\it What} is there---what `substance' are we dealing with?''; a
``dynamical part'', answering the question ``{\it How} does this substance
behave?'';  and a ``phenomenological part'', answering the question ``How
does this substance which is there {\it manifest itself} in a way
accessible to us?''.

In the case of General Relativity, the Kinematics comprises a
{\it differentiable manifold} $M$ of dimension four, a {\it Lorentzian 
metric} $g_{ab}$ on $M$, and a structure which, although it is closely
intertwined with the metric, I want to regard as distinct, namely the
{\it causal order-relation} $\prec$.  The Dynamics is then simply the
Einstein equation $G_{ab} = T_{ab}$, or in case non-gravitational
matter is absent, the purely geometrical statement that the metric is
Ricci-flat.  Finally, the Kinematics and Dynamics manifest themselves
as the familiar Phenomena of length, time, inertia, gravity (in
the narrow sense of the word), causality (for example the impossibility
of signaling faster than light), etc.


 \bigskip
 
{\noindent\bf III. Why quantize? (And what does this mean?)}
\medskip\nobreak     

According to classical General Relativity, the metric behaves
deterministically, but of course this is inconsistent with the stochastic,
quantum behavior of the matter to which the metric couples via the Einstein
equation.  Thus, a theory of quantum gravity in the broadest sense of those
words would just be some theory having both classical gravity and quantum
field theory in flat spacetime as limits (the latter being our best theory
of non-gravitational matter to date).  However, most of us who speak of
{\it quantum} gravity, I think mean something more specific than just this;
and so a major question whose answer defines one's approach to quantum
gravity is:
\item{$\bullet$}{ In what sense do we expect this theory to be {\it quantum}?}

\noindent
To this I would add two further basic questions:
\item{$\bullet$}{Do we need a new kinematics}
 (as well as the new dynamics which ``quantization'' entails)?; and
\item{$\bullet$}{What is the phenomenology of quantum gravity?}

\noindent
The rest of this talk is essentially an essay in answering these three
questions, beginning with the first of them.

\bigskip     
 

{\noindent\bf IV. What is a quantum theory?  (two views)}
\medskip\nobreak          

There are of course many different viewpoints on how quantum mechanics is
to be interpreted, but I will concentrate here on two broadly opposed
attitudes, which I will call the $\Psi$-framework and the
Sum-over-histories framework.

According to the former view the essence of quantum mechanics resides in
its mathematical structure: a {\it Hilbert space}, an {\it algebra of
operators} to be interpreted physically in terms of {\it measurements}; and
a ``{\it projection postulate}'', which tells us how to take the results of
measurements into account in predicting probabilities for future
measurements.  In this framework, the central object is the state-vector
$\Psi$ (which is why I am calling it the $\Psi$-framework), and the
physical interpretation is made in terms of observables.  (See almost any
textbook, for example [25].)

Closely allied with the $\Psi$-framework is the {\it canonical
quantization} approach to quantum gravity.  Although different variants of
this approach may employ different combinations of the basic dynamical
variables, they all work solely with {\it space} (in the sense of a
spacelike hypersurface), as opposed to spacetime.  (For a review of such
issues see [26] [27], for alternative choices of canonical variables
see [28].)

{}From the Sum-over-histories point of view, quantum mechanics is understood
quite differently, namely as {\it a modified stochastic dynamics
characterized by a non-classical probability-calculus in which alternatives
interfere.} To see the essence of quantum mechanics in this way goes back
at least to Heisenberg's Chicago lectures [29], and of course is
associated most closely with the name of Feynman [30].  Within this
framework, the spacetime {\it history} itself is the central object.  It
exists in the same sense in which a history is taken to exist in classical
physics, and the physical interpretation can thus be made directly in terms
of properties of this history---what John Bell called [31] `beables' (a
word that I always thought was some kind of joke until I realized that he
meant it to be pronounced ``be-ables'')---rather than indirectly in terms
of ``observables''.  (See also [17] for a statement of this view.)

Since the sum-over-histories is by nature a ``spacetime approach'', it
naturally leads to a version of quantum gravity which works with
{\it{}spacetime} as opposed to data on a hypersurface.[6]

In comparing these two attitudes, I think it is fair to say that the
$\Psi$-framework is mathematically better developed (although this applies
less to quantum field theory than it does to quantum mechanics in the
narrow sense), whereas the sum-over-histories framework is, to my mind,
more satisfactory philosophically, because it avoids the positivistic
refusal to contemplate anything besides our generalized sense
perceptions.\footnote{*} 
{Formally, the $\Psi$-framework may be seen as a special case of the
sum-over-histories that arises when the amplitude is formed in a suitably
local manner, allowing states $\Psi(t)$ associated with given moments of
time to be introduced as convenient summaries of the past.}

Now why, aside from its philosophical advantages, do I favor the
sum-over-histories/spacetime approach to quantum gravity over the
$\Psi$-framework/canonical quantization approach?  An important part of the
answer has to do with what has been called ``the problem of time'' [32],
though the plural `problems' would probably be a more appropriate word in
this connection.

One such problem which affects the $\Psi$-framework concerns the temporal
meaning of the ``logical ordering'' required by the projection postulate.
In employing that postulate, one writes the projections in a definite
sequence determined by the order of the observations {\it in time}; but how
can such a rule avoid leading to a vicious circle in a theory in which time
itself is one of the things being ``observed''?  

A second, closely related difficulty concerns the ``frozen formalism'' that
results when one applies the formal rules of canonical quantization to
General Relativity (or to any generally covariant Lagrangian theory).  In
consequence of the Hamiltonian constraints, the ``physical observables''
are necessarily all time-independent (they are what Karel Kucha{\v r}
[27] calls `perennials'), and one seems forced into an attempt to ``fix
the time-gauge'' in order to recover a semblance of spacetime from the
disembodied spacelike hypersurface to which the formalism directly refers.
Not only is such a procedure technically questionable, but it can be
dangerous as well: one can easily smuggle arbitrary answers to important
physical questions into the theory in the guise of a ``gauge choice'', for
example to the question whether collapse to a singularity is inevitable in
``mini-superspace cosmology''.

Finally, in a framework based on ``observables'' rather than ``beables'',
how are we to speak about (say) the early universe, if there were no
observers then and none in the offing for a long time to come?  Since some
of the most important applications of quantum gravity are likely to be
precisely to the early universe, this also appears to present a serious
difficulty.

None of these ``problems of time'' would seem to exist for the
sum-over-histo\-ries/space\-time approach.  Time itself doesn't need
to be recovered, because it is there from the very beginning as an
aspect of the spacetime metric.  The projection postulate is
irrelevant, because there is no state-vector to be ``reduced''.  And
the early universe existed just as much as we ourselves do here and
now, even if from our vantage point it is relatively remote and
inaccessible.

There is another point I want to mention here, which is presented more
fully in my contribution to Dieter Brill's Festschrift [33], and that is
that---independently of any problems related to general covariance---the
$\Psi$-framework starts to break down, in a certain sense, already for
quantum field theory in flat spacetime.  One of the seeming advantages of
the $\Psi$-framework vis-as-vis the sum-over-histories is that it appears
to possess a more comprehensive measurement scheme, telling us what in
principle can be measured (every selfadjoint operator) and prescribing (at
least formally) how to design an interaction-Hamiltonian to effect the
corresponding measurement (see [34]).  In contrast, there exists (so far)
no equally comprehensive theory of measurements within the
sum-over-histories framework.  Now this lack is not the great disadvantage
for the sum-over-histories which it would be for the $\Psi$-framework,
because measurement is not a fundamental notion for the
sum-over-histories.  Nonetheless, the question of what can and cannot be
measured is clearly an important one for any theory.  However, it turns out
that the simple measurement scheme which the $\Psi$-framework 
{\it appears} to
possess is physically viable only for quantum mechanics in the narrow sense
of non-relativistic point-particle mechanics.

The point is that in Quantum Field Theory, it proves inconsistent with
causality to assume that every observable constructed from the field
operators within a given spacetime region $R$ can be measured by operations
confined entirely to $R$; to be able to do so would lead to the possibility
of superluminal signaling.  There is no time here to repeat the argument
[33] in detail, but it considers three spacetime regions $A$, $B$, and $C$
arrayed so that communication is possible from $A$ to $B$ and from $B$ to
$C$, but not from $A$ to $C$.  Specifically one can choose $B$ to be a
thickened spacelike hyperplane, with $A$ and $C$ being spacelike separated
points which are respectively to the past and future of $B$.  Assuming that
arbitrary localized ideal measurements were possible in these regions, the
argument concludes that an experimenter stationed at $A$ could transmit
information to a colleague at $C$ by deciding whether or not to perform a
certain observation, given that both know that a certain other observation
{\it will} be performed in the intervening region $B$.  

Thus, one must reject the assumption of arbitrary localized measurements,
and it becomes a priori unclear, for quantum field theory, which
observables can be measured consistently with causality and which can't.
This would seem to deprive the $\Psi$-framework for quantum field theory of
any definite measurement theory, leaving the issue of what can actually be
measured to (at best) a case-by-case analysis, just as it remains (so far)
within the sum-over-histories framework.  (Notice as well that most of the
hypersurface observables with which a canonical formulation of gravity
would presumably deal, are likely to run into locality troubles of this
same sort.)  As pointed out above, this actually puts the $\Psi$-framework
at a disadvantage, because for it, the notion of measurement {\it is}
fundamental.

Finally, I want to return to the more general comparison of the two opposed
frameworks in order to stress what seems to me to be the great practical
advantage of a spacetime approach vis a vis a purely spatial one.  In fact
the questions one wants to ask of a quantum gravity theory are---due
ultimately to the diffeomorphism invariance of General Relativity---all of
an unavoidably spacetime character.  For example, one may want to study how
the horizon area of a black hole responds to the emission of Hawking
radiation.  Or one may want to ask whether the cosmological expansion we
are now experiencing was actually preceded by (say) nine previous cycles of
expansion and recontraction.  Both of these questions make perfect sense if
one has access to an entire 4-geometry, but could one formulate them in
terms of the kind of hypersurface data with which the canonical approach
works?

Well, if we do go in the direction indicated by the signpost reading
``sum-over-histories'', we come a little way down the road to a secondary
fork concerning the proper interpretation of the ``quantum probabilities''
which that formalism yields.  In fact, the sum-over-histories will furnish
a probabilistic answer to almost any question about the history you care to
ask (more precisely, it will furnish relative probabilities for the
elements of any partition of the set of all histories into mutually
exclusive and exhaustive subsets [6]), but it is easy to see on
physical grounds that most of these ``quantum probabilities'' cannot be
very meaningful.  A central question for this approach is therefore, under
what circumstances such probabilities do acquire meaning (an issue
symbolized in my initial laundry list by the question whether probability
is relative or absolute).

Here there are two alternative points of view that I know of.  According to
the first view [35], probabilities are {\it absolute} and
{\it{}unconditional} in the sense that their application rests on no
assumption about the history other than a choice of a cosmological initial
condition, but they have meaning only in the context of a fixed partition
of history space\footnote{*}
{In this sense, it is misleading to describe such probabilities as
 absolute: they are in fact relative to a choice of partition.}
which obeys the condition that Jim Hartle and Murray Gell-Mann call
`decoherence'.  In general however, there are very many decohering
partitions, not all of whose probability-assignments are compatible with
each other, cf. [36].  (An interesting example of such an
incompatibility is that a given partition can have its decoherence
destroyed by subsequent ``observational activity'' which has the effect of
making the original alternatives interfere; or stated more generally and
precisely: there can exist pairs of partitions $P'$ and $P''$, based
respectively on earlier and later properties of the history, such that $P'$
and $P''$ each decohere, but their ``union'' $P'\vee P''$ does not.  Here
$P'\vee P''$ is the partition which asks about both the earlier and the
later properties together; its elements are the atoms of the lattice of
sets generated by the elements of $P'$ and $P''$ via intersection and
complementation.)  According to the second view [37], probabilities are
{\it relative} to a split of the universe into subsystems, and {\it
conditional} on possible assumptions about the behavior (or even existence)
of these subsystems (cf. [38]).  (For example one subsystem could be an
electron and the other a collection of molecules in a cloud chamber.)  The
criterion for the ``quantum probabilities'' to be meaningful is then not
that they necessarily decohere (and hence obey the sum-rules proper to
classical probabilities), but that a sufficiently perfect {\it correlation}
obtain between the two\footnote{*}
{In the meantime, this criterion has evolved.  It now seems that twofold
correlations are not the whole story, but threefold ones might be [39].
Also, with respect to terminology, Jim Hartle has convinced me that the
word ``probability'' should be reserved for a measure that obeys the
standard ``Kolmogorov'' sum-rule; so I would now say ``quantum measure''
instead of ``quantum probability''.  Finally, I think one could improve on
the word ``relative'' used above in connection with the split of the
universe into subsystems.  Its meaning is not (as it might seem to be) that
each possible split carries its own notion of probability, but rather that
prediction on the basis of the quantum measure is possible only in relation
to such splits.}
subsystems (for example the correlation by which the
track in the cloud chamber reflects the path of the electron)
[37].

Incidentally, in arguing for the sum-over-histories/spacetime approach in
preference to the $\Psi$-framework/canonical quantization one, I would not
want to give the impression that I think that mathematical studies of the
operator constraints are necessarily irrelevant.  Indeed, some formulations
of the sum-over-histories effectively employ something akin to a Hilbert
space norm on wave-functions as a mathematical intermediary in computing
quantum probabilities, and making such a formulation mathematically
well-defined might still require a Hermitian inner product on the space of
solutions $\psi$ to the operator-constraints.  On the other hand, a new
kinematics in general {\it would} render the constraints irrelevant (except
possibly in some approximate effective theory), and this brings us to the
second basic question, the one about changing the kinematics.

\bigskip     

\vbox{
{\noindent\bf V. A Modified Kinematics?}
\medskip\noindent
{(a)} {\it Lorentzian or Euclidean Signature?}
}
\smallskip\nobreak

People have suggested several possible modifications of the kinematics of
classical General Relativity, and the one I want to discuss first is
perhaps the least radical---though it's radical enough.  It proposes
[2] that the sum-over-geometries be conducted with positive-definite
(Riemannian) metrics instead of Lorentzian ones.  Such a replacement
carries less fundamental import if one interprets this sum within the
$\Psi$-framework, where the histories have meaning only as intermediaries
used to find a wave-function $\Psi$; but even so, it represents a
significant modification.

The main motivation for altering the signature in this way seems to be that
tunneling phenomena appear, in an ``instanton'' approximation, to occur via
Riemannian solutions of the field equations.  But something like this is
already true in quantum {\it mechanics}, where one can use an
imaginary-time path to compute the WKB approximation to barrier
penetration, as in the classic problem of alpha-decay.  Such a
calculational technique can be interpreted as an infinite-dimensional
saddle point approximation to the path-integral, and from this point of
view the complex-time path, or saddle point, merely {\it summarizes} (to
leading order in $\hbar$) the contribution of a large number of real-time
paths.  Certainly its use would not normally be taken to imply that
physical time turns imaginary while the alpha particle is ``under the
barrier''.  In the same way, a gravitational instanton should presumably be
understood as summarizing the contribution of a large number of Lorentzian
histories.

On this view, the notorious ambiguities [40] in the choice of saddle
point and contour which affect the so-called Euclidean functional integral
will only be resolved (to the extent that this can be done at all without
recourse to an underlying discrete theory) when one has succeeded in
deriving the Euclidean-signature expression by analytic continuation from a
Lorentzian starting point.  Carefully observing (the appropriate infinite
dimensional generalizations of) the rules for deformation of complex
contours ought then to answer such questions as which saddle points
contribute, and what are the relative signs of their contributions.  For my
own enlightenment, I actually went through the corresponding analysis in
detail in the much simpler case of one-dimensional barrier penetration, and
I can affirm that everything works out just as it should, including the
fact that passage through the barrier results in damping rather than
amplification.\footnote{*}
{A sketch of the analysis may be found in the Appendix.}

\medskip\noindent
{(b)} {\it Should We Sum over Different Topologies?}
\smallskip\nobreak

A kinematic question of another type is whether one should include more
than one topology in the sum-over-histories.  (Here, incidentally, I have
posed the question in terms of the sum-over-histories, not only because
that was the ``fork'' we followed earlier, but mainly because the
$3+1$-framework does not lend itself to a dynamical topology in any known
manner.)  To this question, my answer would be: ``yes, and again yes''.
Yes first of all, because it seems contrary to the spirit of Relativity to
make of the topology the only absolute element, which ``affects without
being affected''.

And yes secondly, because of a further reason which is perhaps more
substantive, though not as widely appreciated.  Namely, I want to claim
that the study of topological geons (e.g. [41]) leads to the conclusion
that {\it a dynamical metric requires a dynamical topology}.  The basis of
this claim is the observation that, for a type of particle such as a geon,
whose existence is {\it defined} in terms of the spatial topology, any
procedure of ``second quantization'' by definition forces the topology to
change because the number of geons cannot vary if the topology remains
constant.  But experience shows that a relativistic particle which is only
``first quantized'' is not physically consistent (No matter how you slice
things, you seem inevitably to meet with one or more of the following
difficulties: negative energies, negative probability densities,
non-conservation of probability in scattering, faster than light motion by
the particle (the problem mentioned by Bob Wald yesterday [42]),
inability to measure the particle's energy [43].), and there is no
reason to expect geons to be immune from this imperative.  In this sense, I
believe one can say without too much exaggeration that quantum gravity
without topology change is simply inconsistent physically [44]
[45].\footnote{*}
{This reasoning cannot be airtight, of course, because there might in
theory be an absolute, ``god-given'' topology of spacetime for which no
geons at all can exist, for example $S^3 \times R$.}

A further consideration is that, without topology change, it is possible to
(formally) quantize General Relativity so that certain geons violate the
normal connection between spin and statistics [45] (for a
not-only-formal treatment in $2+1$-dimensions see [46]).  Considering
that some process of pair-creation/annihi\-la\-tion seems always to
underlie the known proofs of spin-statistics theorems, it is thus a natural
guess that the correct incorporation of topology-change into quantum
gravity would automatically set up a correlation between exchange and
rotation which would exclude the spin-statistics violating possibilities
[41].  Some recent evidence in favor of this idea comes from the finding
by Fay Dowker that precisely such a correlation occurs for geons of certain
types which have been pair created via ``U-tube'' cobordisms (this sort of
cobordism being a universal mechanism of pair creation) [47].  If it is
true that a correct incorporation of topology-change must reinstate the
spin-statistics correlation for topological geons, then conversely, the
requirement that this in fact occur will serve as an important test of any
theory of quantum gravity, or to put it another way, as an important clue
in the formulation of such a theory.

\noindent{\underbar {\it Remark}} 
Topology-change not only speaks against canonical quantization, but it also
excludes what is sometimes called the ``covariant'' approach to quantum
gravity, which works with an operator-valued metric field on a {\it fixed}
spacetime manifold [48].

If we do accept that topology-change must be provided for, then we come
still further down the road upon a fork corresponding to my earlier question
whether it is the regularity of the metric or causality that should be
sacrificed, the choice being forced upon us by a theorem of Geroch [49]
according to which topology change entails either singularities or closed
timelike curves.

The choice of allowing closed timelike curves in order to preserve
regularity of the metric can be opposed on the grounds that it conflicts
with the causal set idea, which I will come back to in a moment.  More to
the point, recent evidence is that closed timelike curves (``time loops'')
lead to trouble with the quantum fields living in such a (background)
spacetime, specifically divergences in the stress-energy [5] and a
breakdown of unitarity [50].  (See also Stanley Deser's paper in these
proceedings [51].)  An older, if lesser known difficulty concerns the
pair-creation of monopoles in (5-dimensional) Kaluza-Klein theory.  It
turns out that such pair creation via a globally regular Lorentzian metric
is impossible for topological reasons {\it even if} one allows time loops
to occur (as long as time-{\it orientability} is maintained) [52].  To
me, this strongly indicates that the imposition of strict global regularity
on the metric is inappropriate, a view which is reinforced by considering
how rough the metric is likely to be in any case, given the indications
from quantum mechanics and quantum field theory.

In fact there does exist a mechanism of topology-change which preserves
causality, at the cost of allowing the metric to degenerate to zero at
isolated spacetime points [44] [53].  For any compact cobordism
(interpolating manifold between two spacelike hypersurfaces of possibly
different topology) one can find a metric which degenerates to zero at a
finite number of points, in one of a finite number of predetermined ways
(the number depending on the spacetime dimension), and for which no
time-loops are present.  With respect to such a metric, one can view the
topology change as ``happening'' at the points of degeneration, and one
might accordingly hope for simple dynamical rules to describe what takes
place at such points.  In this direction, there is an intriguing result in
the two-dimensional case, where there are only two possible types of
``elementary cobordism'', corresponding to the ``crotch-point'' in a
``trousers'' spacetime, and the ``crown-point'' in a ``yarmulke'' or ``big
bang'' spacetime.  At the corresponding points of degeneration, the
scalar-curvature Action becomes complex in such a way as to suppress the
former type of topology change and enhance the latter [44] [54].

\medskip\noindent
{(c)} {\it Is the Metric Fundamental, or Only an Effective
           Description of Something Deeper? (Causal Sets)}
\smallskip\nobreak

A much more basic kinematical question than those dealt with so far is
whether the spacetime metric should be replaced by some ``deeper''
structure of which it is only an approximate description.  Now, String
Theory [55] proposes one answer to this, but the answer I want to
discuss follows from still another question: is spacetime ultimately
continuous or discrete?

Here, I cannot resist quoting Einstein, who wrote in 1954,
\smallskip
\singlespace
{\narrower\noindent
     ``The alternative continuum-discontinuum seems to me to be a real
       alternative; i.e., there is no compromise {\dots}
       In a [discontinuum] theory space and time cannot occur {\dots}
       It will be especially difficult to derive something like a
       spatio-temporal quasi-order (!) from such a schema {\dots}
       But I hold it entirely possible
       that the development will lead there \dots '' [56] \smallskip}
\sesquispace
\noindent
(In this quotation the exclamation point is mine, put there because the
words `spatio-temporal quasi-order' seem so obviously to be calling for a
theory based on causal sets!)  Referring to the argument against the
continuum, Einstein goes on to say ``{\dots} This objection is not decisive
only because one doesn't know, in the contemporary state of mathematics, in
what way the demand for freedom from singularity (in the continuum theory)
limits the manifold of solutions.''  Here, the objection was that quantum
mechanics teaches that a bounded system can be described by a finite set of
``quantum numbers'', and such a description  conflicts with the infinite
number of degrees of freedom posited by a continuum theory.  (The loophole
referred to was the possibility that excluding singular solutions of the
field equations might suppress these unwanted degrees of freedom (and
reproduce all the characteristic quantum effects as well, all without
leaving the domain of classical field theory)).

In addition to this argument for a fundamental discreteness there are
several contradictions in existing theories which speak powerfully for
the same conclusion.  These contradictions, which I call ``the three
infinities'' (or perhaps four depending on how you count them), include
the divergences of Quantum Field Theory, the singularities of
classical General Relativity, the apparent non-renormalizability of
naively quantized gravity [57], and the apparently infinite value of the
black hole entropy if no cutoff is present.  The final item in this
list rests on an interpretation of horizon entropy to which I will
return below; to my mind it is the least adequately appreciated of the
common arguments for discreteness.

If we accept all these indications, then we come immediately upon a
subsequent multiple fork in the road corresponding to the question of what
the discrete substratum actually is.  To attempt an answer at this point
would seem to be hopeless if there is not at least what I would call some
sort of ``structural bridge'' between the continuum and the underlying
discontinuum, i.e. some structural analogy which would allow one to
understand how the former can ``emerge'' from the latter in appropriate
circumstances.  People have sought the source of such an analogy in at
least three properties of the continuum, its {\it topology} [58], its
{\it metric} [10] (cf. [59]), and its causal {\it order} [60]; and
I personally have at one time or another been drawn to all of them before
deciding, in connection with the causal set hypothesis, that it is the
{\it{}order} or ``causal structure'' of the spacetime continuum which,
together with one component of the metric (effectively its determinant),
should be viewed as being its most fundamental property.

The causal set hypothesis which I have just alluded to, posits that
the structure of the discrete substratum is that of a locally-finite
partial-ordering (= causal set), and establishes the correspondence
between this underlying structure and the overlying, ``emergent''
Lorentzian manifold by making the causal order and volume-measure of
the latter correspond to the intrinsic order and ``counting measure''
of the former (so spacetime volume = number of elements).

For a general introduction to causal sets and a partial review of work on
that idea see [23].  I will not discuss the subject further here, except to
point to one last fork in the ensuing road which raises the possibility of
allowing the analog of closed timelike curves to occur in the underlying
discrete ordered set.  Such a generalized structure (a ``directed graph'')
would be a possible alternative to the causal set as presently defined, but
it seems to me to be unnatural, because {\it even if} one does let such
``cycles'' occur in the substratum, it still seems impossible to broaden
the rules of correspondence with the continuum in such a way as to allow a
Lorentzian manifold having time loops to emerge as a valid approximation to
a discrete directed graph.  In this sense, one can {\it predict} that time
loops must be absent from the continuum, whether or not their analog is
admitted into the underlying ordered set (cf. [61]).\footnote{*}
{Even if at bottom, spacetime is discontinuous, this of course does not
mean that every continuum theory of quantum gravity is necessarily useless,
since such a theory might still apply at some intermediate level of
approximation.  If so, then one can anticipate that, from the point of view
of the continuum theory, the cutoff coming from the discreteness would
provide a regularizer, and also that an appeal to the discrete theory would
serve to resolve the ambiguities of the continuum theory connected, in
particular, with the effects of nontrivial spacetime topology
(cf. [62]).  Specifically, the deeper theory should be able to
provide the rules that govern processes in which the topology changes.
\vskip -9pt}

\bigskip
 
\vbox{\noindent\raggedright
{\bf VI. A final dynamical fork: should we constrain the four-volume in
the sum over geometries (``unimodular gravity'')?}}
\nobreak     

The alternative ``discrete versus continuous'' was my last one concerning
kinematics; but before turning to phenomenology, I want to raise one
further dynamical question, which is more naturally discussed here than
earlier.  (I say ``further'' because the whole discussion of section IV
was, of course, about dynamics.)  Specifically, the question is whether one
should hold the spacetime volume fixed in the gravitational
sum-over-histories [6].  Unlike with topology-change, the phrasing
of this question in sum-over-histories form is not a matter of principle;
there is an equivalent formulation in canonical terms [63].

Now classically, fixing the volume before varying the Action makes
essentially no difference; its only effect is to convert the
cosmological constant from a free parameter in the Lagrangian into a
free constant of integration of the resulting field equations.  The
physical significance of this ``unimodular'' constraint is therefore
solely quantum mechanical.  The motivation for adopting it comes, in
my mind, first of all from causal set theory, where a constraint on
the total number of elements seems necessary for the sum over causal
sets to converge, and this constraint translates in the continuum into
specifying the total spacetime volume.  Also, independently of any
 discreteness, a direct constraint on the spacetime volume seems
to ameliorate convergence problems with the continuum functional
integral, as is especially noticeable in the limit 
corresponding to quantum 
field theory in curved spacetime.  Most intriguingly, the manner in which
the cosmological constant ``becomes dynamical'' in unimodular quantum
gravity offers a new ``mechanism'' for producing a small or zero value
for it.

It is therefore interesting to ask what difference the unimodular
constraint would make in rudimentary models like the homogeneous universes
of ``mini-super\-space quantum cosmology''.  Recently Jorma Louko and I
have studied a couple of the simplest of such models with results that are
peculiar enough to be interesting, but not so crazy as to become boring
[64].  We find, specifically for the Friedmann universe $S^3\times
\Reals$, that adopting the analog for the unimodular theory of the
``no-boundary boundary-condition'', and computing the crudest saddle point
approximation to the wave-function $\psi$ (with the most obvious saddle
point), that $\psi$ remains regular as a function of $a$, the radius of the
universe, in both the limits $a \to 0$, and $a \to \infty$.  (This is an
example of the improved convergence I spoke of, since in ``standard quantum
cosmology'', $\psi$ diverges exponentially as $a\to\infty$.)  On
the other hand, $\psi$ is now a function of the 4-volume, which serves as a
kind of parameter-time $T$, and its ``evolution'' with $T$ is non-unitary
due to a flux of probability coming in from $a=0$.  One might interpret
this effect as a ``continuous creation of universes'', or perhaps better,
as an ``induced emission of new branches of the universe, all stemming from
a common root''.\footnote{*}
{Or maybe this is just the wrong saddle point.  In an example of the
ambiguity referred to earlier, Jorma has recently found a less obvious
saddle point whose contribution leads to a $\psi$ which is even better
behaved as $a \to \infty$ but which dies out with $T$ instead of blowing
up, suggesting either unitary evolution or a flow of probability {\it out}
through $a=0$!}

\bigskip
 
{\noindent\bf VII. What is the phenomenology of Quantum Gravity?}
\smallskip\nobreak

The third and last set of ``forks'' I want to discuss concerns the
phenomenology of quantum gravity, or more prosaically, the question of
what observable consequences we might expect a theory of quantum
gravity to possess.  If in fact a new substance underlying the
metrical field is the proper basis for such a theory, it becomes
especially important to try to foresee how this new form of matter
will manifest itself (or has already done so!); but even if only the
dynamics of gravity is to be changed, one would expect some dramatic
consequences to appear.  In this connection, let me present a second
laundry list of ``phenomenological'' questions whose answers some
people have hoped would emerge from quantum gravity.

\item{}  Why is there a metric?  And why is it Lorentzian?
     
\item{}  Why is Minkowski space a solution of the theory (with d=4)?  
Notice that this question includes also the question of why the
cosmological constant is so small, since for $\Lambda\ne 0$, Minkowski
space would not be a solution.
     
\item{} Why is the gravitational Lagrangian what it is?
     
\item{} Why does non-gravitational ``matter'' exist (fields and/or particles)?
     
\item{} What is origin of black hole entropy?
     
\item{}    Why is the universe expanding?
     
\item{}    Is CPT broken?
     
\item{}  What are the rules for topology change?

Unlike for my first laundry list, there will be time to address only a
minority of these topics here, and I want to concentrate on two of
them which also occurred in the earlier list, namely those concerning
black hole entropy and the cosmological constant.  Before getting to
them, however, let me allude to part of the answer that causal set
theory would give to the first question of why the spacetime metric is
Lorentzian.  The point is that no other metric signature, Riemannian
or $(++--)$ or whatever, can lend a partial ordering to the events of
spacetime, because only in the Lorentzian case do the light-cones
provide a well-defined local distinction between before and after.

\medskip

\nobreak
(a) {\it Is a Black Hole's Entropy Inside its Horizon?}
\smallskip\nobreak

     Although one might initially think that the entropy of a black
hole must represent the number of its interior states, such a view is
difficult to reconcile with the fact that the second law of
thermodynamics pertains effectively to processes which proceed in
ignorance of whatever is happening inside the horizon.  Since it is by
definition the autonomously developing ensemble of such exterior
processes which are responsible for the entropy increase, it would
seem most natural that the entropy itself be a property of the
exterior region.  In fact, if one adopts this view, and more
particularly if one identifies the exterior entropy with
$S = - {\rm Tr} \, \rho \, \log \rho$, where $\rho$ is the effective
Schr\"odinger-picture density-operator of a spacelike hypersurface in
the exterior region, then there exists a schematic explanation of why
$S$ as so defined necessarily increases as the hypersurface to
which $\rho$ refers advances in time.\footnote{*}
{In using this language I am presupposing, for example, a sufficiently
 classical approximate spacetime with respect to which a given hypersurface
 can be meaningfully located.}
This explanation [65] rests on a certain theorem [66] concerning
density-matrix evolution, and on the crucial fact (emphasized yesterday by
Jimmy York [67]) that the the total energy not only is conserved, but is
meaningful as a property of the exterior region, since it can be read off
from the behavior of the metric in the asymptotic region or on some
suitable boundary surface.

What is more, one can estimate the contribution to the above $S$ from the
zero-point fluctuations of a free scalar field in a black hole background,
and one obtains a value which is proportional to the horizon area measured
in units of the cutoff [68].  It is essentially this result to which I
referred earlier in adducing the black-hole entropy as one of the ``three
infinities''.  Although the story is really more complicated than this
(because the free-field approximation is probably wrong, and it is most
likely the degrees of freedom of the horizon itself which account for its
entropy) I believe that the conclusion that finite entropy requires a
cutoff is correct (cf.  Einstein's objection against the continuum quoted
earlier), and that the sketch of an explanation cited above for the
increase of the total entropy is fundamentally correct also.  If so, then
filling in the sketch so as to obtain a complete derivation of the
increasing character of a well-defined total entropy which includes a
horizon contribution of the correct magnitude will be a decisive test for
any theory of quantum gravity.

\medskip

\nobreak
{(b)} {\it Is the Cosmological Constant Exactly Zero?}
\smallskip\nobreak

I want to conclude with a ``prediction'' about the cosmological
constant, $\Lambda$, which draws together a few of the ideas advanced
so far.  From unimodular gravity let us take the idea that $\Lambda$
is in some sense conjugate to the spacetime volume $V$ (earlier called
`$T$'), and from causal set theory the idea that $V$ is a measure of the
number of elements $N$.  From the the former idea, we can write in the
sense of the Uncertainty Principle,
$$
     \Delta V \, \Delta \Lambda \sim 1.
$$
But since the correspondence between $V$ and $N$ has a probabilistic
character, Poisson fluctuations in $N$ of order of magnitude $\sqrt N$
translate into an uncertainty in $V$ of
$$ 
   \Delta V \sim \Delta N \sim \sqrt N \sim \sqrt V.
$$
Putting these two relationships together yields a minimum uncertainty
in $\Lambda$ of
$$
     \Delta  \Lambda  \sim  {1 \over  \sqrt V},
$$
which, for the visible universe to date, is in order of magnitude,
$10^{-120}$ in natural units.  The prediction [23] is thus that whatever
mechanism drives $\Lambda$ to vanish\footnote{*}
{One possible mechanism is that only $\Lambda=0$ is stable against
non-manifold fluctuations of the causal set.},
will probably leave it with a small but non-zero value of this magnitude,
which, intriguingly, is just barely large enough to be accessible to
observation.

If this is correct then to test a prediction of Quantum Gravity, we might
want to look outward rather than inward, and [69] we might even have the
experimental answers in time for them to be presented at the next birthday
celebration for Dieter and Charlie.

Finally, I would like to thank several colleagues who attended this talk
for their stimulating questions and comments, and Sumati Surya for help
with the references.

\bigskip\noindent
This research was partly supported by NSF grant PHY 9600620.

\bigskip
{\centerline{\bf Appendix: Tunneling and Analytic Continuation}}
\smallskip\nobreak

In this appendix, I sketch the derivation of the WKB barrier penetration
formula from the path integral.  The purpose is not to derive the result as
such, which of course is thoroughly well known, but only to expose the
manner in which imaginary-time paths enter: not as fundamental integration
variables, but only as saddle points of an analytically continued real-time
path integral.  By implication, the use of Euclidean signature metrics in
quantum gravitational calculations is no more reason to doubt the
Lorentzian signature of spacetime than the phenomenon of $\alpha$-decay is
reason to conclude that the physical time of non-relativistic quantum
mechanics is really purely imaginary.

Let us imagine a source of frequency $E$ that emits a particle at one end
of a potential barrier (say at $x=a$) and a sink of the opposite frequency
that absorbs it at the other end (say at $x=b$).  The amplitude for
propagation from emission to absorption is then
$$
    e^{ i S_0(\gamma) - i E \Delta{t}(\gamma)} ,
$$
where
$$
    S_0(\gamma) = \int_\gamma  {m\over 2}{dx^2 \over dt} - V(x) dt
$$
and $\gamma$ is a spacetime path that spends time $\Delta{t}$ going from
source to sink.  The overall amplitude $A$ is thus
$$
        A = \int d\mu(\gamma) \, e^{iS(\gamma)} \ ,              \eqno(1)
$$
where $d\mu(\gamma)$ is the ``measure factor'',
$S(\gamma)=S_0(\gamma)-E\Delta{t}(\gamma)$, and the integral is over all
spacetime paths $\gamma$ that run (with $dt>0$) from $(t,x)=(t_a,a)$ to
$(t,x)=(t_b,b)$.  Notice that $\Delta{t}=t_b-t_a$ is to be integrated over
in (1), unlike the spatial endpoints $a$ and $b$.

It is the path integral (1) that we wish to approximate.  In doing so, I
will follow the usual practice of supposing that manipulations that would
be correct for finite dimensional integrals will also be valid here.  Now
as is common with such problems, we can either continue the integration
``contour'' or continue some parameters in the integrand itself.  Here it
seems clearer to do the latter by introducing a complex parameter $\zeta$
into the action integral as follows:
$$
  S \to S_\zeta 
  = \int_\gamma {m dx^2 \over 2\zeta dt} + (E-V) \zeta dt
$$
(which formally is the same as the substitution $dt\to\zeta dt$).

Now the integral (1) is an oscillating ``Fresnel integral''.  In order
that it remain convergent as $\zeta$ is varied, it is necessary that the
contribution to $iS$
$$
     \int_\gamma {im dx^2 \over 2\zeta dt}
$$
be negative definite.  This implies (since $dt>0$) that $\zeta$ can be
continued freely into the lower half-plane, but not the upper.  Let us
continue it from $\zeta=1$ to the negative imaginary axis and write there
$\zeta = -ic$, with $c>0$.  The action integral then becomes
$$
  iS = \int {-m\over 2c}{dx^2\over dt} - (V-E) c dt .     \eqno(2)
$$
Unlike the original, this integrand has a saddle point (a maximum) within
the domain of integration.  Since $iS$ is now real, we can perform a
steepest descent approximation to $A$ and (noting that $dt$ can be varied
freely, since $\Delta{t}$ is not fixed) we find easily that (ignoring the
prefactor)
$$
        A \sim e^{-I} ,          \eqno(3)
$$
where
$$
     I = \int\limits_a^b dx \sqrt{2m(V-E)} \ .    \eqno(4)
$$
Moreover, since this is independent of $c=i\zeta$, the analytic
continuation back to $\zeta=1$ is trivial and yields exactly the same
answer (4), the familiar WKB result.

Now we have obtained $I$ as the action of a path that extremizes the {\it
analytically continued} action integral (2), a path that proceeds in
``Lorentzian'' time and belongs to the original integration domain of
(1).  But the integral (4) can also be interpreted (for $c=1$)
as the {\it original} action integral $S_{\zeta=1}$ evaluated at the
complex saddle point $\gamma_E$, where $\gamma_E$ is a path running along
the positive imaginary axis in the complex $t$-plane; this interpretation
would have resulted if we had chosen to deform the integration {\it
contour} in (1) rather than the complex {\it parameter} $\zeta$.  But no
matter which way we interpret (4), the right hand side of (3)
is first and foremost an approximation to the ``Lorentzian'' integral
(1).  As such, it represents the sum of the amplitudes of all possible
real-time paths from $a$ to $b$, which, since none of them is a classical
solution at energy $E$, interfere destructively, thereby suppressing the
tunneling.  From the sum-over-histories point of view, the tunneling
particle follows one of these real-time trajectories through the barrier,
although it is impossible to say which one it will be in any particular
case.  A complex-time path like $\gamma_E$, on the other hand, is not a
possible history of the tunneling particle at all, but simply a
mathematical device to help us express the superposition (1) more
compactly.


\vskip 0.5truein
\centerline {\bf References}
\nobreak
\medskip
\noindent
\parindent=0pt
\parskip=10pt
\baselineskip=12pt   



[1] 			
  H.~Minkowski,                        
  ``Space and Time'',
  in {\it The Principle of Relativity},
  translated by W.~Perrett and G.B.~Jeffrey
  (Methuen, 1923, reprinted by Dover).


[2] 			
  A.M.~Polyakov,
  ``Two-dimensional Quantum Gravity. Superconductivity at High $T_c$''
   in E. Br\'ezin and J. Zinn-Justin (eds)
   {\it Fields, Strings and Critical Phenomena},
   Les Houches, Session XLIX, 1988, pp.305-368
   (North Holland 1990).
  
[3] 		
  A.D~Sakharov,
   ``Cosmological transitions with changes in the signature of the metric'',
   {\it JETP \bf 60}:214 (1984); a similar suggestion was made to me after
   my lecture by the honoree, Dieter Brill, who asked whether the Principle
   of Relativity didn't tell us to admit all possible signatures, including
   for example $(++--)$, just as it tells us to admit all possible topologies.

[4]                  
  G.T.~Horowitz,
  ``Topology change in classical and quantum gravity'',
   {\it Class. Quant. Grav.} {\bf 8}: 587-601 (1991).

[5] 			
   K.S.~Thorne,
   ``Closed Timelike Curves'',
      in {\it General Relativity and Gravitation 1992}, 
      Proceedings of the GR13 conference, held Cordoba, 
      Argentina 28 June--4 July, 1992
      (IOP Publishing, Bristol, 1993)
      pp. 295--315.

[6]			
 R.D.~Sorkin, 
``On the Role of Time in the Sum-over-histories Framework for Gravity'',
    paper presented to the conference on 
    The History of Modern Gauge Theories, 
    held Logan, Utah, July 1987, 
    published in  
      {\it Int. J. Theor. Phys.} {\bf 33}:523-534 (1994)

[7] 			
 C.~Teitelboim, ``Proper-time gauge in the quantum theory of gravitation'',
  {\it Phys. Rev. D} {\bf 28}: 297-309  (1983).

[8] 		
  Riemann, G.F.B., 
    {\it  \" Uber die Hypothesen, welche der Geometrie zugrunde liegen,} 
    edited by H. Weyl (Spring\-er-Verlag, Berlin, 1919).

[9]			
 A.A.~Robb,
 {\it Geometry of Time and Space},
 (Cambridge University Press, 1936).

[10] 		
  C.J.~Isham, Y.~Kubyshin and P.~Renteln,
   ``Quantum Norm Theory and the Quantization of Metric Topology'',
    {\it  Class.Quant.Grav.} {\bf 7}:1053 (1990).

[11] 		
 Isham, C.J.
  ``An introduction to general topology and quantum topology'', 
    in 
     {\it Physics, Geometry and Topology}, 
      proceedings of the 
       Advanced Summer Institute on Physics, Geometry and Topology,  
	held  Banff August 14-26, 1989,
	 ed H.C.Lee, pp.129--190 
	 (Plenum Press, New York 1990).

[12]		
 Geroch, R., 
 ``Einstein Algebras'', 
  {\it Comm. Math. Phys.} {\bf 26}:271-275 (1972); the same attitude
  is in effect taken by superstring theorists. 

[13] 		
  S.W.~Hawking,
  ``Space-time Foam'',
    {\it Nucl. Phys.} {\bf B144}:349-362 (1978).
    

[14] 		
   A.~Anderson and B.S.~DeWitt,			
   ``Does the Topology of Space Fluctuate?'',
   {\it Found. Phys} {\bf 16}:91--105 (1986).

[15]			
  R.D.~Sorkin, 
 ``On the Entropy of the Vacuum Outside a Horizon'',
   in B. Bertotti, F. de Felice, Pascolini, A., (eds.),
   {\it Tenth International Conference on General Relativity and Gravitation
   (held Padova, 4-9 July, 1983), Contributed Papers}, 
   vol. II, pp. 734-736
   (Roma, Consiglio Nazionale Delle Ricerche, 1983)

[16] 		
  J.D.~Bekenstein,
   ``Statistical Black Hole Thermodynamics'',
   {\it Phys.~Rev. D} {\bf 12}:3077-3085 (1975).

[17]			
 Sukanya Sinha and Rafael D.~Sorkin,
  ``A Sum-over-histories Account of an EPR(B) Experiment'',
     {\it Found. of Phys. Lett.} {\bf 4}:303-335 (1991).

[18] 		
    R.~Penrose,
   ``Gravity and quantum mechanics'',
     in {\it General Relativity and Gravitation 1992}, 
     Proc. of the GR13 conference, held Cordoba, 
     Argentina 28 June--4 July, 1992
     (IOP Publishing, Bristol, 1993)
     pp. 179--189.
    %

[19] 		
  S.W.~Hawking,
    ``The path-integral approach to quantum gravity'',
    in S.W. Hawking and W. Israel (eds.),
    {\it General Relativity: an Einstein Centenary Survey}
    (Cambridge University Press, 1979), pp. 746--789.

[20] 		
 P.A.M.~Dirac,
  ``The theory of gravitation in Hamilton\-ian form'',
   {\it Proc.~Roy. Soc. London} {\bf A246}:333--343 (1958).

[21]	                
   D.~Finkelstein,  ``Finite Physics''
   in {\it The Universal Turing Machine--A Half-Century Survey },
   R. Herken (ed.), (Hamburg: Kammerer \& Unverzagt, 1987), see
   especially the section on the ``three relativities''.

[22]			
 J.B.~Hartle,
 ``Spacetime Quantum Mechanics and the Quantum Mechanics of Spacetime'',
 in B.~Julia and J.~Zinn-Justin (eds.),
 {\it Les Houches, session LVII, 1992, Gravitation and Quantizations}
 (Elsevier Science B.V. 1995).

[23]			
R.D.~Sorkin, 
 ``Spacetime and Causal Sets'', 
   in J.C. D'Olivo, E. Nahmad-Achar, M. Rosenbaum, M.P. Ryan, 
       L.F. Urrutia and F. Zertuche (eds.), 
   {\it Relativity and Gravitation:  Classical and Quantum,} 
   (Proceedings of the {\it SILARG VII Conference}, 
    held Cocoyoc, Mexico, December, 1990), pages 150-173,
   (World Scientific, Singapore, 1991), and references therein.

[24] 		
  W.~Pauli,
  {\it Theory of Relativity},
  translated from the German by G.~Field
  (Pergamon 1958), Supplementary Note 19, pp.~219--220.
   %


[25]			
    R.~Shankar, 
   {\it Principles of Quantum Mechanics} 
   (Plenum 1980) 
    Chapter 4.

[26] 
  C.J.~Isham,
   ``Conceptual and Geometrical Problems in Canonical Quantum Gravity'',
    in H.~Mitter and H.~Gausterer (eds.),
    {\it Recent Aspects of Quantum Fields}
    (Springer, 1992).	

[27]        		
   K.V.~Kucha\v{r},
   ``Canonical Quantum Gravity''  
       in R.J.~Gleiser, C.N.~Kozameh, O.M.~Moreschi (eds.),
       {\it ``General Relativity and Gravitation 1992 '': Proceedings of the
        Thirteenth Conference on General Relativity and Gravitation}, 
        held Huerta Grande, Cordoba, 28 June-4 July, 1992
        (Bristol:IOP Publishing 1993).

[28]		
    Amitabha Sen, 
 ``Gravity as a Spin System'',
   {\it Physics Letters} {\bf 119B}:89-91 (1982);
   A.~Ashtekar,
   {\it Lectures on Non-perturbative Canonical Gravity},
   (World Scientific 1991);
   C.~Rovelli and L.~Smolin,
    ``Loop representation for quantum General Relativity'',
     {\it Nuc. Phys.} {\bf B}133:80 (1990).

[29]			
  W. Heisenberg,
 {\it The Physical Principles of the Quantum Theory}
  (translated by C. Eckart and F.C. Hoyt)
   (Dover 1930)

[30]			
 R.P. Feynman,
 ``Spacetime approach to non-relativistic quantum mechan\-ics'' 
   {\it Reviews of Modern Physics} {\bf 20:} 367-387 (1948);
  R.P.~Feynman, R.B.~Leighton and M.~Sands,
    {\it The Feynman Lectures on Physics, vol. III: Quantum Mechanics}
    (Addison--Wesley, 1965).

[31]			
 J.S.~Bell,
 {\it Speakable and unspeakable in quantum mechanics: collected papers on
 quantum philosophy}   
 (Cambridge University Press, 1987).

[32] 		
  C.J.~Isham,
  ``Canonical Quantum Gravity and the Problem of Time'',
    in L.~A. Ibort and M.~A.~Rodriguez (eds.), 
    {\it Integrable Systems, Quantum Groups, and Quantum Field Theories}
    (Kluwer Academic Publishers, London, 1993)
     pp. 157--288
    $\langle$e-print archive: gr-qc/9210011$\rangle$
  K.V.~Kucha\v{r},
  ``Time and Interpretations of Quantum Gravity'',
    in G.~Kunstatter, D.~Vincent and J.~Williams (eds.),
    {\it Proceedings of the 4$^{th}$ Canadian Conference on General Relativity
      and Relativistic Astrophysics}
    (World Scientific, 1992).

[33]			
R.D.~Sorkin,
``Impossible Measurements on Quantum Fields'',
   in Bei-Lok Hu and T.A.~Jacobson (eds.),
   {\it Directions in General Relativity: 
     Proceedings of the 1993 International Symposium, Maryland, 
     Vol.~2: Papers in honor of Dieter Brill}, pages 293-305 
   (Cambridge University Press, 1993)
   \eprint{gr-qc/9302018}

[34]			
  J. von Neumann
 {\it Mathematical Foundation of Quantum Mechanics}
  (translated by R.T. Beyer)
  (Princeton U. Press 1955), chapter VI.

[35] 
   See reference [22].  For a still more ``absolutist'' possibility see:
   David Bohm,
   {\it Phys.~Rev. D} {\bf 85}:166 and {\bf 85}:180 (1952).

[36]
 F.~Dowker and A.~Kent, 
 ``On the Consistent Histories Approach to Quantum Mechanics'',
 {\it J. Stat. Phys.} {\bf 82}:1575-1646 (1996)
 \eprint{gr-qc/9412067}

[37]			
R.D.~Sorkin,
``Quantum Mechanics as Quantum Measure Theory'',
   {Mod. Phys. Lett. A} {\bf 9}:3119-3127 (No.~33) (1994)
   \eprint{gr-qc/9401003}

[38]			
  Don~N.~Page,
 ``Interpreting the Density Matrix of the Universe'',
   in A. Ashtekar and J. Stachel (eds.), 
  {\it Conceptual Problems of Quantum Gravity} 
   (Proceedings of the conference of the same name, 
       held Osgood Hill, Mass., May 1988), 116-121
   (Boston, Birkh\"auser, 1991).

[39]
  R.D.~Sorkin,
``Quantum Measure Theory and its Interpretation'', in
    D.H.~Feng and B-L~Hu (eds.), 
    {\it Proceedings of the Fourth Drexel Symposium on Quantum
       Nonintegrability: Quantum Classical Correspondence},
       held Philadelphia, September 8-11, 1994, pages 205--227
    (International Press, 1996)
    \eprint{gr-qc/9507057}.

[40]			
 Jonathan~J.~Halliwell and  J.B.~ Hartle,
``Integration contours for the no-boundary wave function of the universe'',
 {\it Phys Rev D}{\bf 41}:1815 (1990).

[41]			
 R.D.~Sorkin, 
``Classical Topology and Quantum Phases: Quantum Geons'',
   in S.~de~Filippo, M.~Marinaro, G.~Marmo (eds.), 
   {\it Geometrical and Algebraic Aspects of Nonlinear Field Theories}
   (Proceedings of the conference of the same name, 
      held Amalfi, Italy, May 1988), 201-218, 
   (Elsevier, Amsterdam, 1989)

[42] 		
   R.M.~Wald, 
   ``A Proposal for Solving the `Problem of Time' 
     in Canonical Quantum Gravity'',
     talk presented at the
     {\it International Symposium on Directions in General Relativity in
     celebration of the sixtieth birthdays of Dieter Brill and Charles
     Misner}, held College Park Maryland, May 27-29 (1993).

[43]			
  R.D.~Sorkin, 
   ``On the Failure of the Time-Energy Uncertainty Principle'', 
   {\it Found. Phys.} {\bf 9}:123-128 (1979).

[44]			
 R.D.~Sorkin, 
  ``Consequences of Spacetime Topology'', 
     in {\it Proceedings of the Third Canadian Conference on General
          Relativity and Relativistic Astrophysics}, 
           (Victoria, Can\-ada, May 1989), 
             edited  by A. Coley, F. Cooperstock and B. Tupper, 137-163
               (World Scientific, 1990).

[45] 		
 Ch.~Aneziris, A.P.~Balachandran, M.~Bourdeau, S.~Jo, T.R.~Ramadas 
      and R.D.~Sorkin, 
 ``Aspects of Spin and Statistics in Generally Covariant Theories'', 
   {\it Int. J. Mod. Phys.} {\bf A4}, 5459-5510 (1989).

[46] 		
  J.~Samuel,
   ``Fractional Spin from Gravity'', 
   {\it Phys.~Rev.~Lett.} {\bf 71}:215 (1993).

[47]			
 H.F.~Dowker and R.D.~Sorkin,
``A Spin-Statistics Theorem for Certain Topological Geons'',
  (submitted)
  \eprint{gr-qc/9609064}.

[48] 		
  B.~DeWitt,
  ``Quantum Theory of Gravity II: the Manifestly Covariant Theory'',
  {\it Phys.~Rev.} {\bf 162}:1195-1239 (1967);
 S. Mandelstam,
 ``Feynman Rules for the Gravitational Field from the Coordinate-Independent
   Field Theoretic Formalism'',
   {\it Phys. Rev.} {\bf 175}:1604 (1968);           
 N.~Nakanishi and I.~Ojima,                       
   {\it Covariant Operator Formalism of Gauge Theories and Quantum Gravity}
   (World Scientific, Singapore 1990).

[49]			
R.~Geroch, 
 ``Topology in General Relativity'', 
    {\it J.~Math.~Phys.} {\bf 8}:782-786 (1967).

[50]			
  J.L.~Friedman, N.J.~Papastamatiou and J.Z.~Simon, 
 ``Failure of unitarity for interacting fields on spacetimes 
   with closed timelike curves,'' 
   {\it Phys. Rev. D} {\bf 46}:4456 (1992).

[51] 		
  S.~Deser and A.R.~Steif,
  ``No Time Machines from Lightlike Sources in 2+1 Gravity'',
     in Bei-Lok Hu and T.A.~Jacobson (eds.),
     {\it Directions in General Relativity},
       Proceedings of the 1993 Brill-Misner Fest, held Maryland, 
      Vol. 1, p. 78
     (Cambridge University Press, 1993).

[52]			
   R.D. Sorkin, 
 ``On Topology Change and Monopole Creation'',
     {\it Phys. Rev.} {\bf D33}, 978-982 (1986);
  R.D. Sorkin, 
  ``Non-Time-Orientable Lorentzian Cobordism Allows for Pair Creation'', 
      {\it Int. J. Theor. Phys.} {\bf 25}, 877-881 (1986).

[53]		
 Arvind Borde and Rafael~D.~Sorkin,  
 ``Causal Cobordism: Topology Change Without Causal Anomalies'',
   (in preparation).

[54]			
 J.~Louko and R.D.~Sorkin,
``Complex Actions in two-dimensional topology change'',
  {\it Class. Quant. Grav.} {\bf 14}: 179-203 (1997)
  $\langle$e-print archive: gr-qc/9511023$\rangle$.

[55] 		
  M.B.~Green, J.H.~Schwarz and E.~Witten, 
  {\it Superstring Theory}
  (Cambridge University Press, 1987).

[56]			
A.~Einstein, 
 Letter to H.S. Joachim, August 14 1954, Item 13-453,
  cited in J. Stachel,
  ``Einstein and the Quantum: Fifty Years of Struggle'', 
    in {\it From Quarks to Quasars, Philosophical Problems of Modern Physics},
     edited by R.G. Colodny,
      (U. Pittsburgh Press, 1986), pages 380-381.

[57] 		
   See the references in:
     M.Yu~Kalmykov,
    ``Gauge and parametrization dependencies of the one-loop
      counterterms in Einstein gravity''
      {\it Class. Quant. Grav.} {\bf 12}:1401-1412 (1995)
      \eprint{hep-th/9502152}.

[58] 		
  In addition to [11] see
 R.D.~Sorkin, 
 ``A Finitary Substitute for Continuous Topology?'',
   {\it Int.~J.~Theor.~Phys.} {\bf30} 923-947 (1991).
  In effect the so-called ``dynamical triangulation'' approach also makes
  the topology be the fundamental variable (by taking the simplexes of
  ``Regge Calculus'' to be equilateral of unit edge-length, it in effect
  derives a metric from the topology); see
Francois David,
 ``Simplicial Quantum Gravity and Random Lattices'',
 Lectures given at Les Houches, Session LVII, July 5-August 1, 1992:
 Gravitation and Quantizations;
J.~Ambjorn, J.~Jurkiewicz, Y.~Watabiki,
 ``Dynamical Triangulations, a Gateway to Quantum Gravity?'',
 {\it J.~Math.~Phys.} {\bf 36}:6299-6339 (1995)
  $\langle$e-print archive: hep-th/9503108$\rangle$;
 S.~Catterall
  ``Lattice Quantum Gravity: Review and Recent Developments'',
   Talk at international workshop LATTICE 95, held July 1995, Melbourne 
  Australia, 
  {\it Nucl. Phys. B (Proceedings Suppl.)} (in press)
   $\langle$e-print archive: hep-lat/9510008$\rangle$.

[59] 		
   L.M.~Blumenthal and K.~Menger,
   {\it Studies in Geometry}
   (W.H.~Freeman, 1970), 
   Part 3: ``Metric Geometry''.

[60]			
  David Finkelstein,
  `` `Superconducting' Causal Nets'',
  {\it Int. J. Th. Phys} {\bf 27}:473 (1988),
  and references therein;
also see
H.~Reichenbach,
{\it Axiomatik der relativistische Raum-Zeit-Lehre},
translated into English as
{\it Axiomatization of the theory of relativity}
(Berkeley, University of California Press, 1969).

[61]		
  D.R.~Finkelstein,
  {\it Quantum Relativity}
  (Springer 1996).

[62]
R.D.~Sorkin and S.~Surya,
``An Analysis of the Representations of the Mapping Class Group of a 
    Multi-geon Three-manifold'' 
    \eprint{gr-qc/9605050}

[63]			
W.G.~Unruh,
``Unimodular Theory of Canonical Quantum Gravity'',
{\it Phys. Rev.} D {\bf 40}:1048 (1989).

[64]			
A.~Daughton, J.~Louko and R.D.~Sorkin,
``Initial Conditions and Unitarity in Unimodular Quantum Cosmology'',
  in R.B.~Mann and R.G.~McLenaghan (eds.),
    {\it Proceedings of the Fifth Canadian Conference on
      General Relativity and Relativistic Astrophysics} held
           Waterloo, Canada, May, 1993, pp. 181-185
      (World Scientific, 1994)
      $\langle$e-print archive: gr-qc/9305016$\rangle$;
      and a more detailed paper in preparation by the same authors. 

[65]			
 R.D. Sorkin, 
 ``Toward an Explanation of Entropy Increase in the
     Presence of Quantum Black Holes'',
     {\it Phys. Rev. Lett.} {\bf 56}, 1885-1888 (1986).

[66]			
  Woo Ching-Hung,
 ``Linear Stochastic Motions of Physical Systems'',
  Berkeley University Preprint, UCRL-10431 (1962).

[67]			
   J.W.~York,
   "Canonical in all directions---Quasilocal Energy and the
     Microcanonical Functional Integral",
   talk given at the symposium on 
   {\it Directions in General Relativity}, held at the University
   of Maryland, College Park in May, 1993, in honor of Dieter Brill and
   Charles Misner.

[68]			
   R.D.~Sorkin, 
 ``On the Entropy of the Vacuum Outside a Horizon'',
   in B. Bertotti, F. de Felice, Pascolini, A., (eds.),
   {\it Tenth International Conference on General Relativity and Gravitation
   (held Padova, 4-9 July, 1983), Contributed Papers}, 
   vol. II, pp. 734-736
   (Roma, Consiglio Nazionale Delle Ricerche, 1983);
   Bombelli, L., Koul, R.K., Lee, J. and R.D. Sorkin, 
 ``A Quantum Source of Entropy for Black Holes'', 
   {\it Phys. Rev.} {\bf D34}, 373-383 (1986);
  M.~Srednicki, 			
  ``Entropy and Area'',
   {\it Phys.~Rev.~Lett.} {\bf 71}:666-669 (1993)
   $\langle$e-print archive: hep-th/9303048$\rangle$.

[69]			
  Lawrence M.~Krauss and Michael S.~Turner, 
  "The Cosmological Constant is Back",
  \eprint{astro-ph/9504003}.

\end